# Perfect kagome-lattice antiferromagnets with $J_{\text{eff}}$ = 1/2: The Co$^{2+}$-analogs of copper minerals volborthite and vesignieite

Yuya Haraguchi[1,†], Takehiro Ohnoda[1], Akira Matsuo[2], Koichi Kindo[2], and Hiroko Aruga Katori[1]
[1]Department of Applied Physics and Chemical Engineering, Tokyo University of Agriculture and Technology, Koganei, Tokyo 184-8588, Japan
[2]The Institute for Solid State Physics, The University of Tokyo, Kashiwa, Chiba 277-8581, Japan
†chiyuya3@go.tuat.ac.jp

We report the synthesis, crystal structure, and magnetic properties of Co$^{2+}$ kagome magnets Co$_3$V$_2$O$_7$(OH)$_2$·2H$_2$O and BaCo$_3$(VO$_4$)$_2$(OH)$_2$, which can be recognized as Co-analogues of the intensively researched quantum kagome magnet volborthite Cu$_3$V$_2$O$_7$(OH)$_2$·2H$_2$O and vesignieite BaCu$_3$(VO$_4$)$_2$(OH)$_2$. For each compound, the ground state is seemingly A-type antiferromagnetic order. At low temperatures, applying a magnetic field causes a metamagnetic-like transition described by the transition in which antiferromagnetically-aligned canted moments change to ferromagnetically-aligned ones. These ground and field-induced states include a canted ferromagnetic component perpendicular to the kagome planes favored by Dzyaloshinskii-Moriya interactions. These magnetic properties are well characterized by the $J_{\text{eff}}$ = 1/2 physics. Our findings will be the first step toward clarifying the $J_{\text{eff}}$ = 1/2 kagome physics, which has been little studied experimentally or theoretically.

## I. Introduction

Spin-1/2 kagome-lattice antiferromagnets (KAFMs) are one of the most promising spin models to realize a highly entangled quantum spin-liquid (QSL) ground state [1-4]. However, the realistic KAFM materials deviate from the ideal model due to various intertwined perturbations. For example, in herbertsmithite ZnCu$_3$(OH)$_6$Cl$_2$, the "gold standard" in the KAFM exhibiting QSL behavior [5,6], the local spin correlations are greatly perturbed by chemical disorder, the nature of which has not yet been revealed [7-10]. It is also predicted that chemical perturbations will mimic QSL-like behavior [11,12]. Thus, it is crucial to deeply understand the individual roles of perturbations that will always be included in realistic materials. Under this circumstance, the development of new kagome-lattice materials is desirable.

The number of ions that can realize the $S$ = 1/2 state, which can form the kagome-lattice, is minimal. Among them, physicists and material scientists have successively found promising compounds in the mineral database that constitute the kagome-lattice composed of Cu$^{2+}$ ions. Thus, synthetic copper minerals have been the mainstream in quantum frustrated magnetism research. Examples include herbertsmithite, volborthite Cu$_3$V$_2$O$_7$(OH)$_2$·2H$_2$O [13-15], vesignieite BaCu$_3$(VO$_4$)$_2$(OH)$_2$ [16], and Cd-kapellasite CdCu$_3$(OH)$_6$(NO$_3$)$_2$·H$_2$O [17,18]. However, Cu$^{2+}$ ions are highly susceptible to symmetry breaking because of their intense Jahn-Teller activity. Therefore, they tend to deviate from the ideal kagome model. On the other hand, Ti$^{3+}$ ions can also realize $S$ = 1/2, and their Jahn-Teller activity is not as

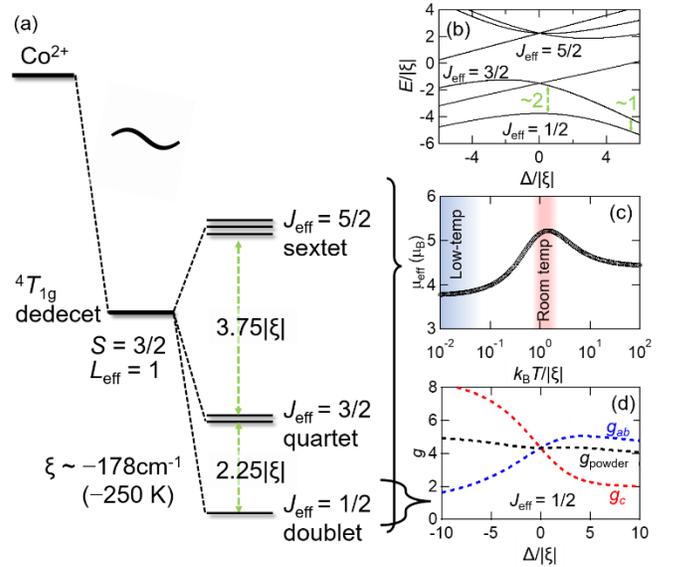

**Fig. 1** (a) Energy levels scheme of high-spin $d^7$ ion under octahedral crystal field and spin-orbit coupling and (b) their trigonal crystal field Δ derivative. (c) The theoretical effective magnetic moment $\mu_{\text{eff}}$ as a function of $k_BT/|\xi|$ in Co$^{2+}$ ion with octahedral crystal fields, where $k_B$ is the Boltzmann constant. (d) The anisotropic and powder-averaged $g$ factors in Co$^{2+}$ ion as a function of $\Delta/|\xi|$.

significant as that of $Cu^{2+}$ ions, so there is a possibility that an ideal kagome model with minor perturbation can be realized. However, all the $Ti^{3+}$ KAFMs reported so far are strongly distorted, resulting in a significant deviation from the kagome model [19,20].

We focus on $Co^{2+}$ as a new candidate ion for achieving a spin-1/2 state. The single-ion magnetic property of $Co^{2+}$ in an octahedral environment is determined by the 12-fold degenerated (dodecet) ground state manifold $^4T_{1g}$ [21-23]. Due to spin-orbit coupling, this dodecet splits into three manifolds—doublet, quartet, and sextet, as shown in Fig. 1(a). The lowest-energy state is given by $J_{eff} = 1/2$ Kramers doublet. As shown in Fig. 1(b), $J_{eff} = 1/2$ is always the lowest energy regardless of the trigonal distortion, and the energy gap from the excited state is approximately $|\xi|$ to $2|\xi|$. Therefore, if the temperature is a much smaller energy scale than the spin-orbit interaction $\xi = -178$ cm$^{-1}$ ($\sim -250$ K), the magnetic properties are governed by the lowest energy $J_{eff} = 1/2$ Kramers doublet.

Therefore, treating $Co^{2+}$ compounds as a $J_{eff} = 1/2$ doublet physics is possible in the low-temperature region. However, physicists must pay carefulness to the temperature dependence in the physical properties. The contribution of the excited states, $J_{eff} = 3/2$ quartet and $J_{eff} = 5/2$ sextet, by the Boltzmann distribution must cause the effective magnetic moment to change with temperature. Figure 1(c) shows the theoretically-calculated $\mu_{eff}$ for non-interacting $Co^{2+}$ in an octahedral crystal field with spin-orbit couplings $H = \xi L \cdot S$, and $L$ is the total orbital quantum number [24]. It is written as

$$\mu_{eff}(Co^{2+}) = \sqrt{\frac{3\left\{\frac{63x}{20}+\frac{98}{25}+\left(\frac{640x}{225}+\frac{4312}{2025}\right)e^{\frac{-15x}{4}}+\left(\frac{169x}{36}-\frac{490}{81}\right)e^{-6x}\right\}}{x\left\{3+2e^{\frac{-15x}{4}}+e^{-6x}\right\}}} \quad (1)$$

, where $x = \xi/k_B T$. This theoretical calculation yields $\mu_{eff} \sim 5.2$ $\mu_B$ near room temperature and $\sim 4$ $\mu_B$ at low temperatures. Conversely, detecting the temperature variation of $\mu_{eff}$ makes it possible to demonstrate a realization of $J_{eff} = 1/2$ physics in realistic $Co^{2+}$ materials at low temperatures.

In addition, physicists should also pay carefulness to the effects of low-symmetry crystal fields when considering only the $J_{eff} = 1/2$ doublet at low temperatures. In a perfect octahedral crystal field, there is no anisotropy in the Zeeman splitting of the doublet in a magnetic field. On the other hand, in an axis-symmetric crystal field such as a trigonal or tetragonal crystal field, there is a change in energy between the case parallel to the c-axis and the case perpendicular to it. Therefore, anisotropy appears in the $g$-value. Figure 1(d) shows the g-values expressed as a function of the parameter $\Delta/|\xi|$, where $\Delta$ is the trigonal distortion energy [22,25]. Here, $\Delta/|\xi|<0$ means trigonally-elongated the $CoO_6$ octahedron along the c-axis and $\Delta/|\xi|>0$ trigonally-compressed. The powder-averaged $g_{powder}$ ($= [(2g_{ab}^2 + g_c^2)/3]^{0.5}$) of the powder sample is almost unchanged and is approximate $g_{powder} \sim 4$, although there is a slight $\Delta$ dependence. Therefore, the experimental observation of a powder-averaged $g$-value close to 4 at low temperatures in the powder sample is a criterion for satisfying the $J_{eff} = 1/2$ condition.

Unlike $Cu^{2+}$ or $Ti^{3+}$ KAFMs, there is a limited number of candidates for the $Co^{2+}$ KAFMs, so experimental studies have been limited. $Co_3Mg(OH)_6Cl_2$ contains $\sim 7\%$ site-mixture between the $Mg^{2+}$ and $Co^{2+}$ sites, resulting in a spin glass character due to randomness [26]. Moreover, synthesis of $BaCo_3(VO_4)_2(OH)_2$ [27] and $Co_3V_2O_7(OH)_2 \cdot 2H_2O$ [28], $Co^{2+}$-analogues of vesignieite and volborthite, have also been reported. Their crystal structures are shown in Figs. 2(a) and 2(b). Both compounds commonly comprised $Co_3O_6(OH)_2$ kagome layer [Figs. 2(c)]. Both compounds exhibit superparamagnetism without magnetic orderings, probably due to finite-size effects [29,30].

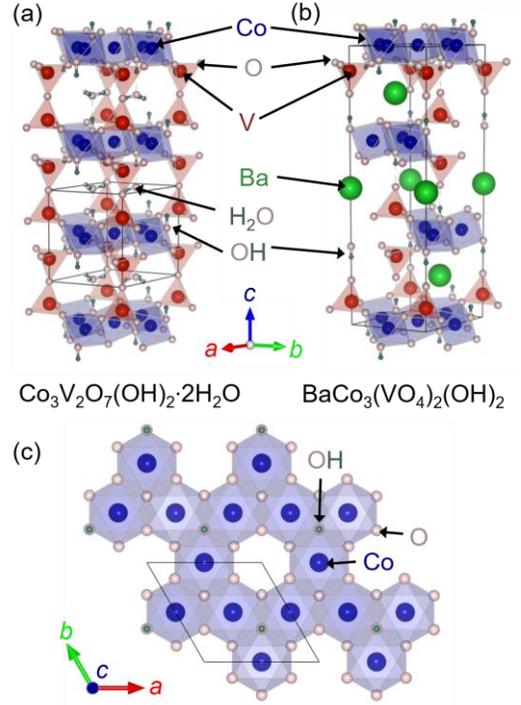

**Fig. 2** Crystal structures of (a) $Co_3V_2O_7(OH)_2 \cdot 2H_2O$ and (b) $BaCo_3(VO_4)_2(OH)_2$ viewed along the $c$ axis. (c) The $Co_3O_6(OH)_2$ kagomé layer viewed along the $ab$ plane commonly observed in $Co_3V_2O_7(OH)_2 \cdot 2H_2O$ and $BaCo_3(VO_4)_2(OH)_2$. The VESTA program is used for visualization [31].

This work reports the successful synthesis of high-quality $Co_3V_2O_7(OH)_2 \cdot 2H_2O$ and $BaCo_3(VO_4)_2(OH)_2$ polycrystalline samples, as well as their magnetic properties. The crystal structure analysis demonstrates the formation of a perfect kagome-lattice without distortion and site mixture. Moreover, the magnetic and thermodynamic properties commonly observed in both materials are well characterized in the $J_{eff} = 1/2$ state. Our findings demonstrate that $Co_3V_2O_7(OH)_2 \cdot 2H_2O$ and $BaCo_3(VO_4)_2(OH)_2$ are promising model compounds $J_{eff} = 1/2$ KAFMs.

## II. Experimental Methods

Polycrystalline samples were prepared using the hydrothermal method. All starting materials were purchased from FUJIFILM Wako Pure Chemical Corporation. For

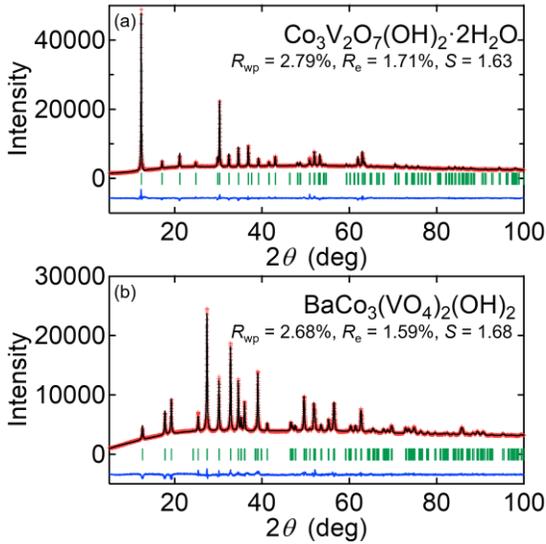

**Fig. 3** (a) Powder x-ray diffraction patterns of (a) $Co_3V_2O_7(OH)_2·2H_2O$ and (b) $BaCo_3(VO_4)_2(OH)_2$. The observed intensities (red), calculated intensities (black), and their differences (blue) are shown. Green vertical bars indicate the positions of Bragg reflections.

$Co_3V_2O_7(OH)_2·2H_2O$, a Teflon beaker containing 0.37g of $Co(NO_3)_2·6(H_2O)$, 0.1 g of $NH_4VO_3$, 0.07 g of NaOH, and 10 ml of pure $H_2O$, was heated at 120ºC for 60 h. For $BaCo_3(VO_4)_2(OH)_2$, a Teflon beaker of 30 ml volume containing 0.35 g of $Ba(OH)_2·8H_2O$, 0.78 g of $CoCl_2·2H_2O$, 0.14 g of $NH_4VO_3$, and 5 ml of pure $H_2O$, was heated at 200ºC for 12 h. In both syntheses, orange-colored powders were obtained after rinsing with distilled water several times and drying at 110 °C. The thus-obtained samples were characterized by powder x-ray diffraction (XRD) using Cu-Kα radiation. The cell parameters and crystal structure were refined by the Rietveld method using the RIETAN-FP version 2.16 software [32]. The temperature dependence of magnetization was measured under magnetic fields up to 7 T in a magnetic property measurement system (MPMS; Quantum Design). The temperature dependence of heat capacity was measured by a conventional relaxation method in a physical property measurement system (PPMS; Quantum Design). Magnetization curves up to approximately 50 T were measured by the induction method in a multilayer pulsed magnet at the International Mega Gauss Science Laboratory in Institute for Solid State Physics.

## III. Results

### A. Crystal structures

Figures 3(a) and (b) show the powder XRD pattern of $Co_3V_2O_7(OH)_2·2H_2O$ and $BaCo_3(VO_4)_2(OH)_2$. For both samples, the observed XRD patterns are perfectly reproduced well using previously-reported crystal structure parameters without changing any parameters, confirming that our samples are single-phase and match the trigonal structures. The crystal structure commonly in $Co_3V_2O_7(OH)_2·2H_2O$ and $BaCo_3(VO_4)_2(OH)_2$ contains an edge-sharing network of $CoO_6$ octahedra as shown in Fig. 2(c), where the Co atoms form a perfect kagome-lattice. There is only-one crystallographic Co site and Co-Co distance in the kagome-lattice. Each Co ion is surrounded by six anions: two OH ions and four O ions. The bonds are short for the OH ions at the trans position, and long for the four lateral O ions. The 2-short-4-long-type crystal field must affect the single-ion anisotropy in the $J_{eff} = 1/2$ pseudospin of $Co^{2+}$.

### B. Magnetic Property

Figures 4(a) and (b) show the temperature dependencies of magnetic susceptibility χ and their inverse 1/χ for $Co_3V_2O_7(OH)_2·2H_2O$ and $BaCo_3(VO_4)_2(OH)_2$. In the 1/χ curves, a slope change is observed around 50 K. The effective magnetic moments $μ_{eff}$ and Weiss temperature $θ_W$ obtained by the Curie-Weiss (CW) fits of the 1/χ curves in high-temperatures (150-300 K) and low-temperatures (10-25 K) are listed in the Table 1. In common with both materials, the $μ_{eff}$ values are ~5.2 $μ_B$ for high-temperature regions and ~4.2 $μ_B$ for low-temperature regions, which is roughly consistent with the theoretical calculation shown in Fig. 1(c). Thus, the $μ_{eff}$-shrinkage is explained in terms of level split owing to spin-orbit couplings. These negative Weiss temperatures indicate that the interaction between $Co^{2+}$ spins is predominantly antiferromagnetic. Furthermore, the change in slope of 1/χ as a function of temperature is observed. For $Co^{2+}$ ions in an octahedral environment, the spacings of the multiplet levels ($J = 1/2, 3/2$, and $5/2$) are not so large compared to $k_BT$. Thus, the observation of $μ_{eff}$-decreasing demonstrates a crossover of the multiplet levels.

In both materials, the χ-data increases and then tends to saturate at low temperatures, a characteristic of ferromagnets. On the contrary, as shown in Figure 4(c), the χ-data of $Co_3V_2O_7(OH)_2·2H_2O$ and $BaCo_3(VO_4)_2(OH)_2$ under lower applied fields (10 mT) exhibit antiferromagnetic-like magnetic anomalies at $T_N$ = 2.7 and 4.9 K, respectively. A clear magnetic order appeared in these materials, whereas no magnetic order was found in previous reports [29,30], probably due to improved crystallinity and sample quality. In addition, the slight separation of the zero-field cooling (ZFC) and field cooling (FC) curves below $T_N$ suggests an emergent magnetic ordering. Figure 4(d) shows the isothermal magnetizations $M$ and their derivative $dM/dH$ at 2 K for $Co_3V_2O_7(OH)_2·2H_2O$ and $BaCo_3(VO_4)_2(OH)_2$. The peaks in the $dM/dH$ data at 75 mT for $Co_3V_2O_7(OH)_2·2H_2O$ and $μ_0H_M$ = 16 mT for $BaCo_3(VO_4)_2(OH)_2$ indicate a magnetic field-induced phase transition. The $M^2$ vs. $H/M$ plot [the inset of Fig 4(e)] shows an "S" like behavior because of first-order metamagnetic transition according to Landau theory [33]. Also, metamagnetism is universally observed in A-type antiferromagnets with intralayer-ferromagnetically-coupled layers [34-38]. The origin of the A-type-antiferromagnetic-ferromagnetic phase transition at low magnetic fields appears to be due to weak interlayer coupling. Therefore, we conclude that the title materials are A-type antiferromagnets.

**Table 1** Experimentally obtained Curie-Weiss parameters at high- (150-300 K) and low-temperatures (10-25 K) in $Co_3V_2O_7(OH)_2·2H_2O$ and $BaCo_3(VO_4)_2(OH)_2$.

| Formula | 150-300 K | | 10-25 K | |
|---|---|---|---|---|
| | $μ_{eff}$ ($μ_B$) | $θ_W$ (K) | $μ_{eff}$ ($μ_B$) | $θ_W$ (K) |
| $Co_3V_2O_7(OH)_2·2H_2O$ | 5.138(3) | −17.4(3) | 4.49(3) | 3.6(1) |
| $BaCo_3(VO_4)_2(OH)_2$ | 5.087(2) | −22.7(2) | 4.25(5) | 1.6(1) |

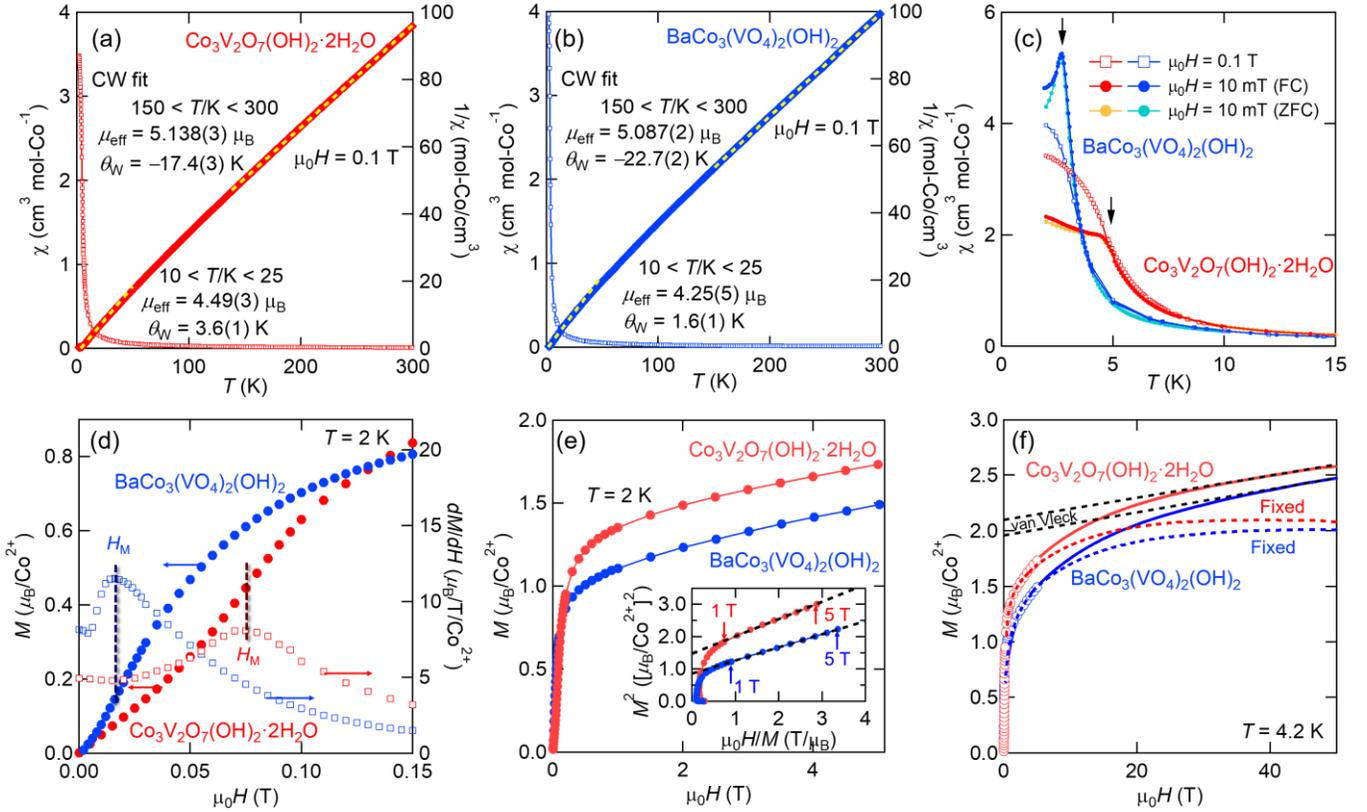

**Fig. 4** Temperature dependence of the magnetic susceptibility $\chi$ (open symbols) and inverse magnetic susceptibility $1/\chi$ (closed symbols) for (a) $Co_3V_2O_7(OH)_2 \cdot 2H_2O$ and (b) $BaCo_3(VO_4)_2(OH)_2$. The dotted lines on the $1/\chi$ data represent fits to the Curie-Weiss (CW) model at high- ($150 < T/K$) and low-temperature ($10 < T/K < 25$) regions. (c) Temperature dependences of $\chi$ measured for magnetic fields of 0.1 T and 10 mT. (d) Low field ($< 0.15$ T) magnetization curves $M$ and their derivative $dM/dH$ measured at 2 K for $Co_3V_2O_7(OH)_2 \cdot 2H_2O$ and $BaCo_3(VO_4)_2(OH)_2$. The metamagnetic-like transition is observed at $\mu_0 H_M = 75$ mT for $Co_3V_2O_7(OH)_2 \cdot 2H_2O$ and $\mu_0 H_M = 16$ mT for $BaCo_3(VO_4)_2(OH)_2$. (e) The $M$ data at 2 K up to 5 T for $Co_3V_2O_7(OH)_2 \cdot 2H_2O$ and $BaCo_3(VO_4)_2(OH)_2$. The inset shows the $M^2$ vs $\mu_0 H/M$ plot. (f) Magnetization curves $M$ measured at 4.2 K under pulsed magnetic fields up to 50 T for $Co_3V_2O_7(OH)_2 \cdot 2H_2O$ and $BaCo_3(VO_4)_2(OH)_2$. The magnitudes are calibrated to the data measured under static fields up to 5 T (open circles).

The good linearity of the $M^2$ vs. $H/M$ plot with a positive value intercept in the 1 ~ 5 T field region indicates that the spontaneous magnetization $M_0$ in the category of Landau theory is an order parameter in the field-induced phase [39]. The $M_0$ values obtained as the intercept of the linear fit are $M_0 = 1.21$ $\mu_B$ for $Co_3V_2O_7(OH)_2 \cdot 2H_2O$ and $M_0 = 0.924$ $\mu_B$ for $BaCo_3(VO_4)_2(OH)_2$. In general, the ground state of $Co^{2+}$ is well described by $J_{eff} = 1/2$ pseudospin as shown in Fig. 1(a), and the powder-averaged g-factor is ~ 4 [see Fig. 1(d)]; thus, $M_{sat}$ ~ $2\mu_B$. Compared to $M_{sat}$, the observed $M_0$ is relatively small. Figure 4(f) shows the isothermal magnetizations of $Co_3V_2O_7(OH)_2 \cdot 2H_2O$ and $BaCo_3(VO_4)_2(OH)_2$ up to 50 T. Above 40 T, the $M$ data increase almost linearly due to the large temperature-independent van Vleck paramagnetism of octahedrally-coordinated $Co^{2+}$ ions. Linear fits the $M$ data above 40 T yield the van Vleck paramagnetic susceptibility $\chi_{vv}$ is $5.57 \times 10^{-3}$ cm³/mol-$Co^{2+}$ for $Co_3V_2O_7(OH)_2 \cdot 2H_2O$ and $5.59 \times 10^{-3}$ cm³/mol-$Co^{2+}$ for $BaCo_3(VO_4)_2(OH)_2$. Then, the saturation magnetizations $M_{sat}$ were obtained as $M_{sat} = 2.10$ $\mu_B/Co^{2+}$ for $Co_3V_2O_7(OH)_2 \cdot 2H_2O$ and $M_{sat} = 1.98$ $\mu_B/Co^{2+}$ for $BaCo_3(VO_4)_2(OH)_2$. Thus, the observed $M_{sat}$ value provides strong evidence that $J_{eff} = 1/2$ ground state is realized in these materials, yielding the respective powder average g-values as follows: $g = 4.20$ in $Co_3V_2O_7(OH)_2 \cdot 2H_2O$ and $g = 3.96$ in $BaCo_3(VO_4)_2(OH)_2$. The observation of $M_{sat}$ ~ 2 demonstrates that the reduced $M_0$ is not due to temperature effects but rather due to the presence of a field-induced phase with a small spontaneous magnetization.

### C. Thermodynamic property

Figure 5(a) shows the temperature dependence of the heat capacity divided by temperature $C/T$ of $Co_3V_2O_7(OH)_2 \cdot 2H_2O$ and $BaCo_3(VO_4)_2(OH)_2$. Clear $\lambda$-shaped peaks are observed at 4.6 and 2.7 K for $Co_3V_2O_7(OH)_2 \cdot 2H_2O$ and $BaCo_3(VO_4)_2(OH)_2$, respectively. Therefore, both in $Co_3V_2O_7(OH)_2 \cdot 2H_2O$ and $BaCo_3(VO_4)_2(OH)_2$, second-order

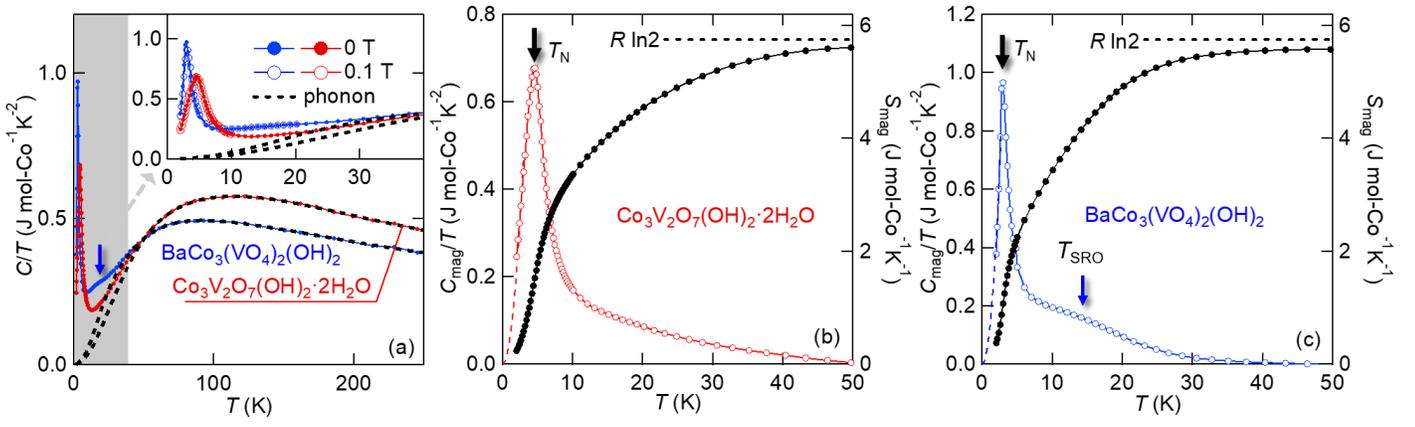

**Fig. 5** (a) The total heat capacities divided by temperature $C/T$ of $Co_3V_2O_7(OH)_2\cdot 2H_2O$ and $BaCo_3(VO_4)_2(OH)_2$. The inset shows the enlarged view of shaded area. The $C/T$ data with 0.1 T magnetic field applied is also shown as open circles. The dashed lines on the data represent the lattice contribution estimated by fitting the data above 100 K, as described in the text. The temperature dependences of the magnetic heat capacity divided by temperature $C_{mag}/T$ after the subtraction of lattice contribution and magnetic entropy $S_{mag}$ for (b) $Co_3V_2O_7(OH)_2\cdot 2H_2O$ and (c) $BaCo_3(VO_4)_2(OH)_2$. The horizontal dashed lines indicate the value of $S_{mag} = R\ln 2$, which is the total magnetic entropy derived from $J_{eff} = 1/2$.

magnetic phase transitions with bulk nature occur at $T_N$. As shown in the inset, even at 0.1 T, where the magnetic anomaly seems to disappear in the magnetic susceptibility, a λ-type peak is almost unchanged. This fact indicates that the magnetic field region above $H_M$ is a robust magnetically ordered state different from the ground state, contrasting with a polarized phase.

Generally, to evaluate the magnetic contribution, the heat capacity of the nonmagnetic analog is used to determine their lattice contributions. Unfortunately, the nonmagnetic counterpart $BaZn_3(VO_4)_2(OH)_2$ is unavailable, and the $C/T$ data of $Zn_3V_2O_7(OH)_2\cdot 2H_2O$ (as a natural mineral martyite [40]) does not match the high-temperature $C/T$ data of $Co_3V_2O_7(OH)_2\cdot 2H_2O$ at all (see the Supplemental Material [41]). Thus, to extract the magnetic contribution to the heat capacity, the lattice contribution was estimated by fitting the high-temperature part where the magnetic heat capacity may be negligible. It is well known that a lattice heat capacity $C_{latt}$ consists of contributions from three acoustic phonon branches and $3n-3$ optical phonon branches, where $n$ is the number of atoms per formula unit [42]; $n$ equals 22 for $Co_3V_2O_7(OH)_2\cdot 2H_2O$ and 18 for $BaCo_3(VO_4)_2(OH)_2$. The acoustic and optical contributions are described by the Debye- and Einstein-type heat capacities $C_D$ and $C_E$, respectively. Provided that $C_{latt}$ is the sum of $C_D$ and $C_E$, the $C/T$ data above 100 K, where the magnetic heat capacity may be negligible, are fitted to the equation,

$$C_{lattice} = C_D + C_E$$
$$= 9R(T/\theta_D)^3 \int_0^{\theta_D/T} \frac{x^4 \exp(x)}{[\exp(x)-1]^2} dx$$
$$+ R \sum_{i=1}^{3} n_i \frac{(\theta_{Ei}/T)^2 \exp(\frac{\theta_{Ei}}{T}-1)}{\exp(\frac{\theta_{Ei}}{T}-1)} \quad (4)$$

, where $R$ is the gas constant, $\theta_D$ is the Debye temperature, $\theta_{Ei}$ is the Einstein temperatures. The best fits are shown by the dashed lines with $\theta_D$ = 180(2) K, $\theta_{E1}$ = 260(9) K, $\theta_{E2}$ = 552(12) K, $\theta_{E3}$ = 1677(24) K, $n_1$ = 17, $n_2$ = 23, and $n_3$ = 23 for $Co_3V_2O_7(OH)_2\cdot 2H_2O$ and $\theta_D$ = 153(4) K, $\theta_{E1}$ = 227(9) K, $\theta_{E2}$ = 633(11) K, $\theta_{E3}$ = 1544(52) K, $n_1$ = 16, $n_2$ = 17, and $n_3$ = 18 for $BaCo_3(VO_4)_2(OH)_2$. Figures 5(b) and 5(c) show the magnetic contribution $C_{mag}/T$ obtained by subtracting this lattice contribution from the experimental data and the magnetic entropy $S_{mag}$ calculated by integrating the magnetic $C_{mag}/T$ with respect to $T$ assuming the $C_{mag}/T$-value equals 0 at 0 K following the third law of thermodynamics. The asymptotic value of $S_{mag}$ at high temperatures close to $R\ln 2$ = 5.76 mol$^{-1}$K$^{-1}$ expected for a doublet demonstrates that the $J_{eff}$ = 1/2 doublet is realized.

As shown in Fig. 5(c), the $C_{mag}/T$ data of $BaCo_3(VO_4)_2(OH)_2$ shows a broad hump at $T_{SRO} \sim 15$ K higher than $T_N$, indicating a short-range ordering. This hump can also be seen in the unsubtracted $C/T$ data [indicated by the blue arrow in Fig. 5(a)]. On the other hand, as shown in Fig. 5(b), the $C_{mag}/T$ data for $Co_3V_2O_7(OH)_2\cdot 2H_2O$ shows no apparent hump. However, up to around 50 K, which is considerably higher than $T_N$, there is a finite $C_{mag}/T$ that decays with increasing temperature, indicating the development of short-range order. Furthermore, the $S_{mag}$ data at $T_N$ reach approximately 20% and 12% for $Co_3V_2O_7(OH)_2\cdot 2H_2O$ and $BaCo_3(VO_4)_2(OH)_2$, respectively. These observations indicate that the development of short-range magnetic correlations above $T_N$ has released magnetic entropy. The difference in the presence/absence of a distinct hump structure in $C_{mag}/T$ data associated with the short-range ordering probably depends on the difference in magnetic dimensionality derived from their crystal structures—the $V_2O_7$ polyanions bound to the kagome layer of $Co_3V_2O_7(OH)_2\cdot 2H_2O$ are tightly chemically bound to the neighboring kagome layer, whereas the $VO_4$ polyanions bound to the kagome layer of $BaCo_3(VO_4)_2(OH)_2$ are isolated from the neighboring kagome layer. Therefore, the magnetic dimensionality must be derived from the "built-in" character of a crystal structure, being consistent with the higher $T_N$ and $H_M$ in $Co_3V_2O_7(OH)_2 \cdot 2H_2O$.

## IV. Discussion

$Co_3V_2O_7(OH)_2 \cdot 2H_2O$ and $BaCo_3(VO_4)_2(OH)_2$ are $Co^{2+}$ analogues of the kagome copper minerals volborthite and vesignieite. The observed effective magnetic moment, saturated magnetization, and magnetic entropy suggest a realization of $J_{eff} = 1/2$ doublet at low temperatures.

First, the observed metamagnetic-like transition with reduced $M_0$ is discussed. Both title materials exhibit $\sim 1/2 M_{sat}$ after metamagnetic-like transition, distinct from conventional metamagnets. Moreover, the maximum magnetization increases without saturation on increasing the applied magnetic field above $H_M$. The observed $M_{sat} \sim 2$ expected for $J_{eff} = 1/2$ probed by pulsed magnetization measurements demonstrates that the small $M_0$ is not due to temperature effects. This behavior is consistent with a noncollinear spin structure such as a canted antiferromagnetic order. In a noncollinear spin structure, spin canting creates an uncompensated net moment. The coplanar 120-degree spin structure is most stable without Dzyaloshinskii-Moriya (DM) interactions in the classical kagome-lattice antiferromagnets [43].

On the other hand, due to finite DM interactions, all spins tilt from the kagome plane to an umbrella-like structure, resulting in a spontaneous magnetization [44-46]. Essentially, there is no inversion center at the middle point of the coupling between neighboring magnetic sites in the kagome-lattice. Such magnetic structures have been universally observed in several jarosite-type materials [47,48] and rare-earth kagome $R_3Sb_3Mg_2O_{14}$ ($R$ = Nd, Dy) [49,50]. Hence, spontaneous magnetization emerges within a single kagome layer. In this situation, a metamagnetic-like transition in which antiferromagnetically-aligned net moments in the layer change to ferromagnetically-aligned ones can emerge, as shown in Fig. 6. Such a metamagnetic-like transition of the net moment has been observed in jarosite-type $AFe_3(OH)_6(SO_4)_2$ ($A$ = K, Ag) [51,52].

Next, we discuss the magnitude of the single-layer net magnetization. The order of the $D$ vector can be estimated using the following equation [53,54],

$$|D| \sim \frac{g - g_{spin}}{g} 2J \quad (5)$$

, where $g_{spin}$ (= 2) is the spin-only $g$-value, and $J$ is the exchange interaction. The canting angle $\theta$ from the plane of coplanar spin structure with the competition and DM interactions as

$$\tan\theta = \frac{|D|}{2J} \sim \frac{g - g_{spin}}{g} . \quad (6)$$

The expected in-plane spontaneous magnetization $M_0$ using the $\theta$-value obtained with the experimentally observed $g \sim 4$ is calculated to be $M_0 \sim 0.45 M_{sat}$. Therefore, it is possible to explain the observed reduced $M_0$ from the DM interaction alone in $Co^{2+}$ ions. Note that the spontaneous magnetization of the field-induced phases is powder-averaged since the measurements were performed on powder samples. Thus, it is difficult to determine whether the reduced $M_0$ originates solely from DM interactions in the present stage. Therefore, the future establishment of single crystal growth will clarify the exact origin of the reduced $M_0$. Moreover, the absence of the 1/3 plateau characteristic of kagome-lattice antiferromagnets would be due to the in-plane net moment of $\sim 1/2 M_{sat}$ in canting spin structure transcending that of $1/3 M_{sat}$ in the up-up-down one. Neutron diffraction experiments are a future issue to clarify the details of these magnetic structures, and it is urgent to establish a synthetic method for deuterated title materials $Co_3V_2O_7(OD)_2 \cdot 2D_2O$ and $BaCo_3(VO_4)_2(OD)_2$.

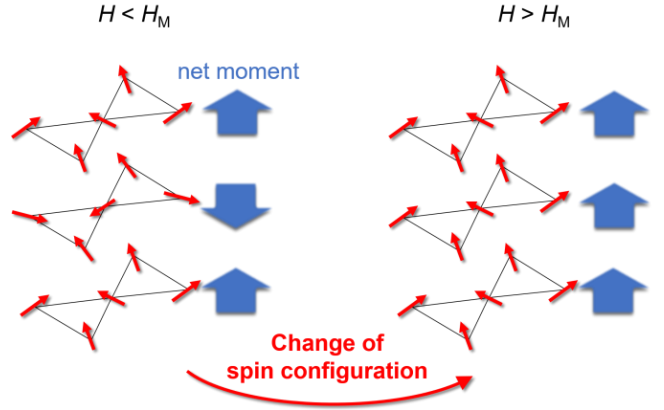

**Fig. 6** Schematic view of the change of spin configuration in which the antiferromagnetically aligned canted moments changes to ferromagnetically aligned ones in $Co_3V_2O_7(OH)_2 \cdot 2H_2O$ and $BaCo_3(VO_4)_2(OH)_2$.

Finally, we discuss the influence of DM interactions on magnetic susceptibility. By Moriya, an extended Curie-Weiss rule that incorporates the DM interaction by molecular field approximation has been given as follows,

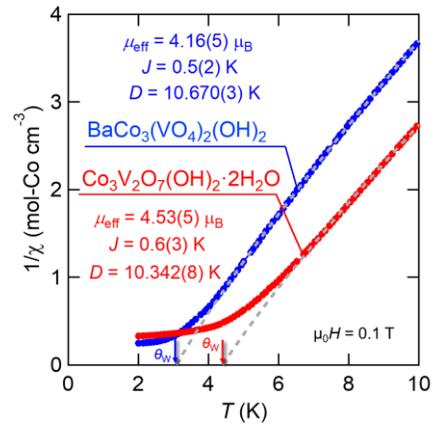

**Fig. 7** Results of the modified Curie-Weiss fit incorporating the Dzyaloshinskii-Moriya interaction in the molecular field approximation to the the $1/\chi$ data on the temperature dependence of the inverse susceptibility at low-temperature ($7 < T/K < 10$) regions.

$$\chi = \frac{N\mu_{\text{eff}}^2}{3k_B(T-\theta_w)}\frac{(T+T_0)}{(T+\theta_w)} + \chi_0, \quad (7)$$

with

$$\theta_w = \frac{JzS(S+1)}{3k_B}\left[1+\left(\frac{|\boldsymbol{D}|}{J}\right)^2\right]^{\frac{1}{2}}, \quad (8)$$

$$T_0 = \frac{JzS(S+1)}{3k_B}, \quad (9)$$

,where $z$ indicates the coordination number; $z = 4$ in kagome lattice [54]. The results of the fitting using Eq. (7) for the $1/\chi$ data below 10 K are shown in Fig. 7. Here, to reduce fitting errors, $\chi_0$ was fixed to the value of $\chi_{vv}$ estimated from the high-field magnetization process [see Fig. 4(f)]. The fits yield $\mu_{\text{eff}} =$ 4.53(5) $\mu_B$, $J = 0.6(3)$ K, and $|\boldsymbol{D}| = 10.342(8)$ K for $Co_3V_2O_7(OH)_2\cdot 2H_2O$, and $\mu_{\text{eff}} = 4.16(5)$ $\mu_B$, $J = 0.5(2)$ K, and $|\boldsymbol{D}| = 10.670(3)$ K for $BaCo_3(VO_4)_2(OH)_2$. The thus-obtained $\theta_W$-values, corresponding to the intercept of the $T$-axis, are 4.4 K and 3.0 K for $Co_3V_2O_7(OH)_2\cdot 2H_2O$ and $BaCo_3(VO_4)_2(OH)_2$, respectively, roughly consistent with the magnetic transition temperatures $T_N$. These results indicate that relatively large DM interactions dominantly stabilize the magnetic order. Here, given the relationship in Eq. (5), it is mysterious that $J$ is so small relative to $|\boldsymbol{D}|$, of which the reasonable origin is a cancellation of several competing superexchange interactions with different signs. For example, ferromagnetic Kitaev interactions expected for $Co^{2+}$ magnets possibly compete with conventional Heisenberg antiferromagnetic interactions [55,56]. In the present stage, extracting them independently from the magnetic susceptibility analysis is impossible. Therefore, inelastic neutron scattering experiments are planned to estimate the exchange parameters needed to construct reasonable spin models for $Co_3V_2O_7(OH)_2\cdot 2H_2O$ and $BaCo_3(VO_4)_2(OH)_2$.

## V. Summary

We have synthesized the high-quality polycrystalline samples of $Co_3V_2O_7(OH)_2\cdot 2H_2O$ and $BaCo_3(VO_4)_2(OH)_2$ via a hydrothermal route and investigated their magnetism and thermodynamic properties. We have shown that they are promising model compounds for the $J_{\text{eff}} = 1/2$ kagome-lattice magnets. The observed canted ferromagnetic state with a halfway-saturated magnetic net moment originates from the DM interaction and the large g-factor, which would be characterized as an emblematic property of the $J_{\text{eff}} = 1/2$ kagome magnets. The $J_{\text{eff}} = 1/2$ kagome magnets will provide us with a unique platform to study quantum magnetism.

## Acknowledgment

This work was supported by Japan Society for the Promotion of Science (JSPS) KAKENHI Grant Number JP22K14002, JP19K14646, and JP21K03441. Part of this work was carried out by joint research in the Institute for Solid State Physics, the University of Tokyo.

---

**Supplemental material for "Perfect kagome-lattice antiferromagnets with $J_{\text{eff}}$ = 1/2: The $Co^{2+}$-analogs of copper minerals volborthite and vesignieite"**

Yuya Haraguchi[1,†], Takehiro Ohnoda[1], Akira Matsuo[2], Koichi Kindo[2], and Hiroko Aruga Katori[1]
[1]Department of Applied Physics and Chemical Engineering, Tokyo University of Agriculture and Technology, Koganei, Tokyo 184-8588, Japan
[2]The Institute for Solid State Physics, The University of Tokyo, Kashiwa, Chiba 277-8581, Japan
†chiyuya3@go.tuat.ac.jp

In order to estimate the lattice contribution of heat capacity in $Co_3V_2O_7(OH)_2\cdot 2H_2O$, we measured the heat of $Zn_3V_2O_7(OH)_2\cdot 2H_2O$ (natural mineral Martyite), in which a nonmagnetic Zn ion substitutes the Co site of $Co_3V_2O_7(OH)_2\cdot 2H_2O$. Figure S1 shows the heat capacity of $Zn_3V_2O_7(OH)_2\cdot 2H_2O$. For comparison, the specific heat data of $Co_3V_2O_7(OH)_2\cdot 2H_2O$ (the same as in the text) and the calculated phonon contribution are also included. The C/T data for both materials do not agree over the entire temperature range, and especially not at all good agreement is obtained in the high-temperature range. The origin of this discrepancy is probably due to the difference in lattice vibrations caused by the different strength of the local chemical bonds between transition metal ions and oxygen ions in Co and Zn, although the two materials are structurally very similar. Hence, $Zn_3V_2O_7(OH)_2\cdot 2H_2O$ is not a suitable nonmagnetic reference material for estimating the lattice-specific heat in $Co_3V_2O_7(OH)_2\cdot 2H_2O$.

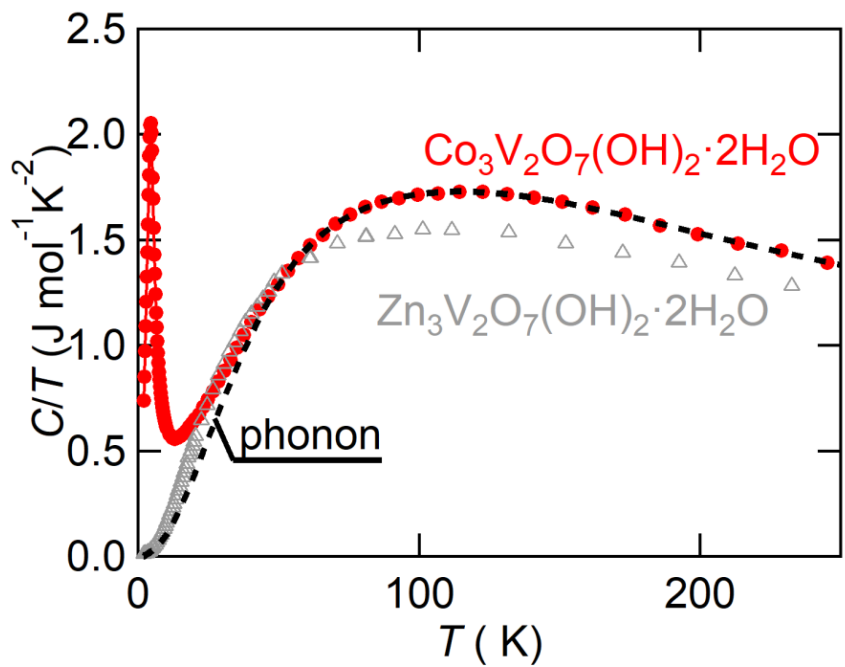

**Fig. S1.** The temperature dependence of $Co_3V_2O_7(OH)_2\cdot 2H_2O$ and $Zn_3V_2O_7(OH)_2\cdot 2H_2O$. The calculated phonon contribution in $Co_3V_2O_7(OH)_2\cdot 2H_2O$ is also displayed (See the main text for calculation method).